\documentclass[journal=cmatex,manuscript=article,layout=twocolumn]{achemso}

\usepackage{siunitx}
\usepackage{amsmath}
\usepackage{amssymb}
\newcommand\ce[1]{\ensuremath{\mathrm{#1}}}
\newcommand\ch[1]{\ensuremath{\mathrm{#1}}}

\usepackage{textcomp}
\usepackage{gensymb}
\usepackage{textgreek} 
\usepackage{color,soul}
\usepackage{wrapfig}
\usepackage{lipsum}
\setkeys{acs}{usetitle=true} 
\SectionNumbersOn

\newcommand\NxMP{Na$_{\rm x}$M$_{2}$(PO$_{4}$)$_{3}$}
\newcommand\NxVP{Na$_{\rm x}$V$_{2}$(PO$_{4}$)$_{3}$}
\newcommand\NxTP{Na$_{\rm x}$Ti$_{2}$(PO$_{4}$)$_{3}$}
\newcommand\NxCP{Na$_{\rm x}$Cr$_{2}$(PO$_{4}$)$_{3}$}
\newcommand\EKRA{$E_{\rm KRA}$}
\newcommand\Ebarrier{$E_{\rm barrier}$}

\title{
Kinetic Monte Carlo Simulations of Sodium Ion Transport in NaSICON Electrodes
}

\let\oldmaketitle\maketitle
\let\maketitle\relax

\author{Ziliang Wang}
\affiliation[National University of Singapore]
{Department of Materials Science and Engineering, National University of
Singapore, 9 Engineering Drive 1, 117575, Singapore}

\author{Tara P.\ Mishra}
\affiliation[National University of Singapore]
{Department of Materials Science and Engineering, National University of
Singapore, 9 Engineering Drive 1, 117575, Singapore}
\altaffiliation{Singapore-MIT Alliance for Research and Technology, 1 CREATE Way, 10-01 CREATE Tower, Singapore 138602, Singapore}

\author{Weihang Xie}
\affiliation[National University of Singapore]
{Department of Materials Science and Engineering, National University of
Singapore, 9 Engineering Drive 1, 117575, Singapore}

\author{Zeyu Deng}
\affiliation[National University of Singapore]
{Department of Materials Science and Engineering, National University of
Singapore, 9 Engineering Drive 1, 117575, Singapore}

\author{Gopalakrishnan Sai Gautam}
\affiliation[IISC]
{Department of Materials Engineering, Indian Institute of Science, Bengaluru, Karnataka, 560012, India}

\author{Anthony K.\ Cheetham}
\affiliation[National University of Singapore]{Department of Materials Science and Engineering, National University of
Singapore, 9 Engineering Drive 1, 117575, Singapore}
\altaffiliation{Materials Department and Materials Research Laboratory, University of California, Santa Barbara 93106 California, USA}

\author{Pieremanuele Canepa}
\affiliation[National University of Singapore]
{Department of Materials Science and Engineering, National University of
Singapore, 9 Engineering Drive 1, 117575, Singapore}
\affiliation{Department of Chemical and Biomolecular Engineering, National University of Singapore, 4 Engineering Drive 4, 117585, Singapore}
\altaffiliation{Department of Electrical \& Computer Engineering, University of Houston, Houston, Texas, 77204, United States of America} 
\email{pcanepa@nus.edu.sg,  pcanepa@central.uh.edu}

\input{insbox}

\begin{document}
\twocolumn[
\begin{@twocolumnfalse}
\oldmaketitle
 \begin{abstract}
 \InsertBoxR{0}{\includegraphics[width=0.6\linewidth]{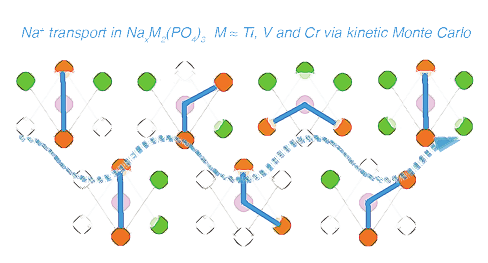}}
\noindent The development of high-performance sodium (Na) ion batteries requires improved electrode materials. The energy and power densities of Na superionic conductor (NaSICON) electrode materials are promising for large-scale energy storage applications. However, several practical issues limit the full utilization of the theoretical energy densities of NaSICON electrodes. A pressing challenge lies in the limited sodium extraction in low Na content NaSICONs, e.g., $\rm Na_1V^{IV}V^{IV}(PO_4)_3  \leftrightarrow  V^{V}V^{IV}(PO_4)_3 + 1e^- + 1Na^+$. Hence, it is important to quantify the  Na-ion mobility in a broad range of NaSICON electrodes. Using a kinetic Monte Carlo approach bearing the accuracy of first-principles calculations, we elucidate the variability of Na-ion transport vs.\ Na content in three important NaSICON electrodes, \NxTP{}, \NxVP{}, and \NxCP{}. Our study suggests that  Na$^+$ transport in NaSICON electrodes is almost entirely determined by the local electrostatic and chemical environment set by the transition metal and the polyanionic scaffold. The competition with the ordering-disordering phenomena of Na-vacancies also plays a role in influencing Na-transport. We identify the Na content providing the highest room-temperature diffusivities in these electrodes, i.e., \ch{Na_{2.7}Ti_2(PO_4)_3}, \ch{Na_{2.9}V_2(PO_4)_3}, and \ch{Na_{2.6}Cr_2(PO_4)_3}. We link the variations in the Na$^+$ kinetic properties by analyzing the competition of ligand field stabilization transition metal ions and their ionic radii.
 We interpret the limited Na-extraction at $x = 1$ observed experimentally by gaining insights into the local Na-vacancy interplay. We propose that targeted chemical substitutions of transition metals disrupting local charge arrangements will be critical to reducing the occurrence of strong Na$^+$-vacancy orderings at low Na concentrations, thus, expanding the accessible capacities of these electrode materials. 

\end{abstract}
\end{@twocolumnfalse}
]
\clearpage

\section{Introduction}

Having achieved widespread commercialization, rechargeable lithium (Li)-ion batteries (LIBs) are now at the risk of geopolitically constrained supply chains of key raw materials, such as cobalt, nickel, and Li.\cite{tarascon_is_2010,olivetti_lithium-ion_2017,turcheniuk_ten_2018} 
Sodium (Na)-ion batteries (SIBs) appear promising alternatives to the LIB analogs, as Na-metal can be harvested directly from seawater\cite{palomares_na-ion_2012,larcher_towards_2015,chayambuka_liion_2020}. Extensive research is underway to optimize electrodes and electrolytes for SIBs.\cite{kaufman_understanding_2019,deng_phase_2020,deb_jmr_2022,singh_chemical_2021,kim_electrode_2012,hasa_challenges_2021,masquelier_polyanionic_2013,wang_phase_2022,chotard_discovery_2015,nogai_dipole_2018,kawai_high-voltage_2018,senguttuvan_low-potential_2013,lalere_improving_2015,lalere_all-solid_2014,zhang_high_2016}  One of the material classes for NIBs is the polyanionic sodium superionic conductor (NaSICON), discovered by \citeauthor{hong_crystal_1976}, \citeauthor{goodenough_fast_1976}\cite{hong_crystal_1976,goodenough_fast_1976}, a framework studied for its fast Na-conducting properties. Electrodes crystallizing in the NaSICON framework, with formula {\ch{Na_x M_2(PO_4)_3}} (where \ch{M} = transition metal) can be highly tuned to achieve promising energy densities,\cite{singh_chemical_2021,wang_phase_2022,park_crystal_2022} by changing the ratio and types of transition metals in the NaSICON, such as \ch{Na_xTiV(PO_4)_3}, \ch{Na_xTiMn(PO_4)_3}, \ch{Na_xVMn(PO_4)_3} and \ch{Na_xCrMn(PO_4)_3}.\cite{masquelier_polyanionic_2013,singh_chemical_2021} 

For most NaSICON electrodes, the accessible capacity   is significantly lower than the theoretical value, which is linked to difficulties in reversibly extracting the whole available Na content. For example, in \NxVP{} the reversible extraction of four sodium ions entails the utilization of all vanadium redox states (\ch{V^{V}/V^{IV}}, \ch{V^{IV}/V^{III}}, and \ch{V^{III}/V^{II}}) accounting for a theoretical gravimetric capacity of $\sim$235~mAh~g$^{-1}$. 
In practice, only two sodium ions can be reversibly extracted from \ch{Na_3V_2(PO_4)_3} up to \ch{Na_1V_2(PO_4)_3}.\cite{chotard_discovery_2015,lalere_coupled_2018,ishado_exploring_2020,wang_phase_2022,park_crystal_2022} 
While \citeauthor{gopalakrishnan_vanadium_1992}  suggested the possibility of chemically extracting the last \ch{Na^+} forming \ch{V_2(PO_4)_3},\cite{gopalakrishnan_vanadium_1992}  successive endeavors have proven unsuccessful. Hence, understanding the factors that limit reversible Na extraction within NaSICONs and facilitating the same remains an active topic of research.

In this letter, using density functional theory (DFT) based kinetic Monte Carlo (kMC) simulations, we unveil the physical origins for the variation in Na$^+$ transport properties in three NaSICON electrode materials, namely {\NxTP}, {\NxVP}, and {\NxCP}. Our analysis reveals the kinetic limits to reversibly extracting \ch{Na}-ions in these NaSICONs apart from identifying the Na-composition ranges to achieve high Na$^+$ diffusivity. The macroscopic Na$^+$ transport is highly influenced by the interplay between Na-vacancy arrangements and transition metals. Our results shed light on the optimization of NaSICON electrodes for improved reversible capacities.

\section{Results}

To investigate the {\ch{Na^+}} transport in \NxTP, \NxVP, and \NxCP{} we rely on a combination of DFT calculations, constructing a cluster expansion Hamiltonian, and performing kinetic Monte Carlo simulations (\emph{vide infra}).\cite{van_der_ven_nondilute_2008,deng_fundamental_2022} The ground-state structures representing specific Na-vacancy arrangements at different Na-compositions of the \NxTP, \NxVP, and \NxCP{}  NaSICON were taken from Refs.~\citenum{singh_chemical_2021} and \citenum{wang_phase_2022}. To estimate the {\ch{Na^+}} migration barriers in the  three NaSICONs, we selected several sodium compositions, including  \ch{x}~=~1, 3, and 4 in \NxTP{}, \ch{x}~=~1, 2, 3, and 4 of \NxVP{}, and \ch{x}~=~1, and 3 of \NxCP{}, respectively. 

At {\ch{x}}~=~1 and 4, all three NaSICONs crystallize in the rhombohedral space-group ($R\overline{3}c$ or $R\overline{3}$).  While {\ch{Na_4Cr_2(PO_4)_3}} is included for completeness in this investigation, it is not expected to be stable (due to the instability of Cr$^{\rm II}$ in the solid state \cite{long_evaluating_2020}  and has never been reported experimentally.   {\ch{Na_3M_2(PO_4)_3}} configurations with monoclinic (\textit{Cc} or \textit{C2/c}) symmetry represent the compositions with the lowest formation energies in the {\ch{Na_xM_2(PO_4)_3}} pseudo-binary tie line of all the three NaSICONs.\cite{singh_chemical_2021,wang_phase_2022} Also, recent investigations predicted the mixed-valence {\ch{Na_2V_2(PO_4)_3}} as a thermodynamically stable phase.\cite{wang_phase_2022,park_crystal_2022}  

Each Na-vacancy (Va) configuration at the above Na compositions was optimized using the strongly constrained and appropriately normed exchange and correlation functional within DFT.\cite{sun_strongly_2015} A Hubbard {\emph{U}} correction (SCAN+\emph{U}) was applied, which has been confirmed to accurately predict the redox electrochemistry during Na (de-)intercalation.{\cite{wang_phase_2022,sai_gautam_evaluating_2018,long_evaluating_2020,devi_effect_2022, deng_fundamental_2022}}  We use the nudged  elastic band (NEB) method to simulate the \ch{Na^+} migration barriers.\cite{sheppard_optimization_2008}

\begin{figure}[!ht]
    \centering
    \includegraphics[width=1\columnwidth]{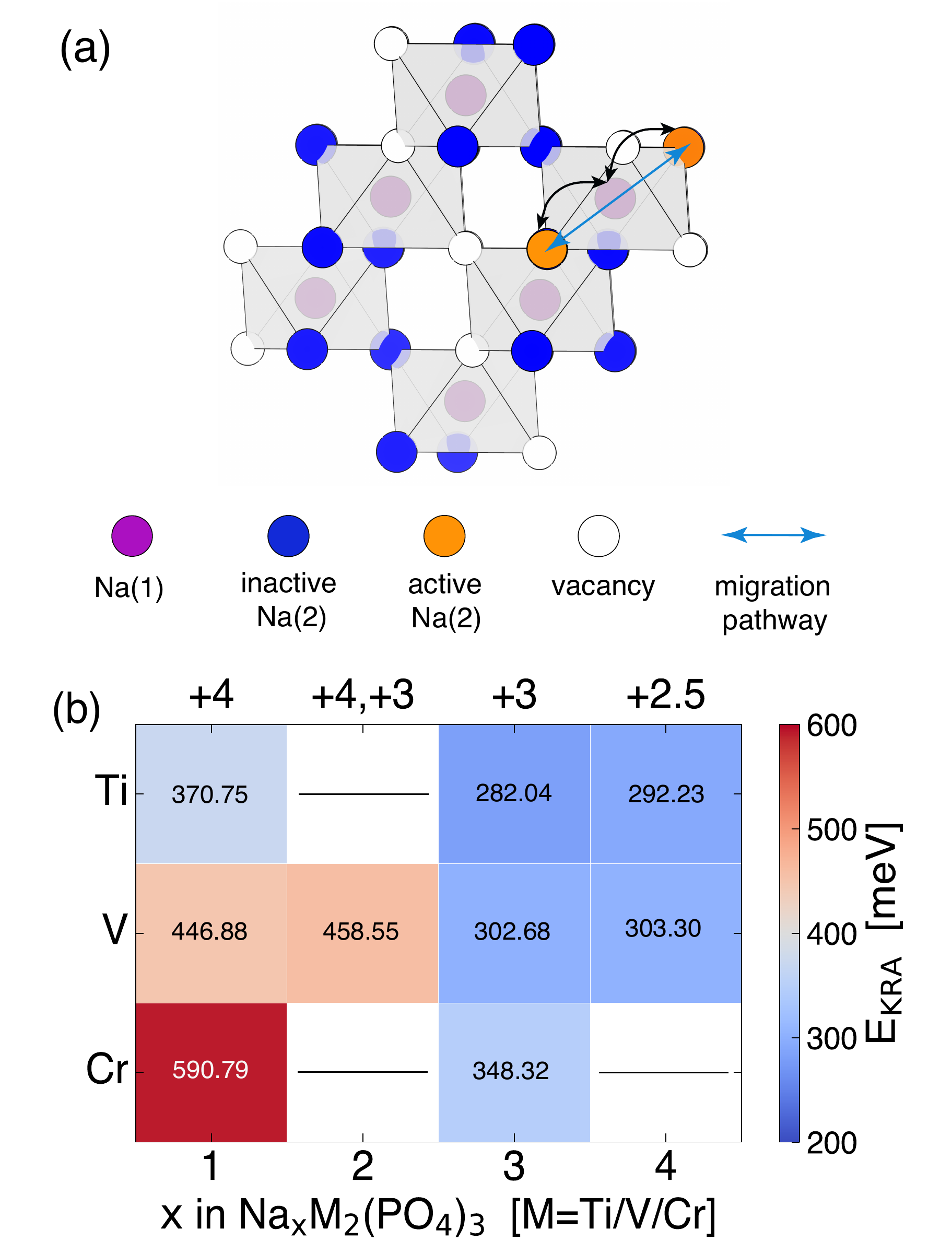}
    \caption{Model of {\ch{Na^+}} migration in the NaSICON \ch{Na_x M_2(PO_4)_3} identifying the migration unit (MU). \textbf{(a)} A representation of the Na-vacancy sublattice in NaSICON, illustrated by the corner-sharing octahedra. Two Wyckoff positions, the Na(1) (6\emph{b}) sites are shown by purple circles, and the \ch{Na}(2) (18\emph{e}) sites by orange or blue circles, depending on the participation of the Na(2) site in a migration event. Empty circles denote vacancies. Each {\ch{Na}}(1) site is surrounded by six nearest neighbor \ch{Na}(2) sites. In each octahedron, two \ch{Na}(2) sites participate in the \ch{Na}-ion migration pathway (orange circles connected by blue line), whereas every single hop  \ch{Na}(2)$\longleftrightarrow$\ch{Na}(1) is denoted by the black double-arrows. \textbf{(b)} shows the computed \EKRA s for the migration event \ch{Na}(2)$\longleftrightarrow$\ch{Na}(1) in the MU (values are indicated in each box), with varied \ch{Na} compositions (x) and transition metals (M). Black lines indicate scenarios where migration barriers were not computed (see main text).  Na compositions are shown in the bottom x-axis, and the oxidation states of transition metals are in the top x-axis.
    }
    \label{fig:Structure_Ekra}
\end{figure}

Typically, the migration mechanism of ions in fast conductors is ascribed to be a local property of the immediate chemical environment of the migrating species, which is strongly influenced by the local ion-vacancy configuration(s).\cite{van_der_ven_first-principles_2001, van_der_ven_nondilute_2008, van_der_ven_rechargeable_2020, deng_fundamental_2022} Figure~\ref{fig:Structure_Ekra} shows the relevant local portion of the NaSICON structure ---the migration unit (MU)--- that is sufficient to capture the Na-migration with variations in the local configurations of Na and vacancies.  Figure~\ref{fig:Structure_Ekra}(a) represents the general MU in \ch{Na_x M_2(PO_4)_3}, where groups of corner-shared octahedra
contain six Na(2) sites centered around a Na(1) site. For the sake of visualization the \ch{MO_6} and the \ch{PO_4^{3-}} groups are not shown. Within each MU octahedron, two Na(2) sites will be ``active'' taking part in the {\ch{Na}}(2)$\longleftrightarrow${\ch{Na}}(1)$\longleftrightarrow${\ch{Na}}(2) migration event, whereas the remaining four Na(2) sites and/or vacancies will be ``inactive'' (Figure~{\ref{fig:Structure_Ekra}}(a)) and spectating the process of Na migration. Herein, we consider a single \ch{Na} migration hop \ch{Na}(2)$\longleftrightarrow$\ch{Na}(1)  as the fundamental Na-ion migration event in a single MU.\cite{goodenough_fast_1976,hong_crystal_1976,boilot_relation_1988, chen_challenges_2017,ishado_exploring_2020, deng_fundamental_2022} 

The collective diffusion of \ch{Na}-ions in the NaSICON electrodes can be captured by our lattice model, which is composed of thousands of MUs, with different {\ch{Na}}-vacancy orderings. Indeed, ensembles of MUs are used to tessellate periodically  the \NxMP{}  structures in the composition range 1~$\le \rm x \le$~4.
At intermediate compositions (1 $< \rm x <$~4), a variety of MUs with different local Na vacancy arrangements is sufficient to approximate the migrating environments 
of the Na ions.  The subsets of crystallographic sites of the migrating Na-ions are fully encompassed by an ensemble of MUs incorporating all possible Na-vacancy occupation arrangements. Therefore, the partial occupations of Na sites as refined from experimental techniques, such as X-ray diffraction,\cite{kabbour_-na_2011,wang_phase_2022,park_crystal_2022,lucazeau_neutron_1986} can be described by the 3D networks formed by thousands of MUs.

Based on the assumption that the transport property relies mainly on the local Na-vacancy orderings, all the possible migration pathways within the MUs were extensively calculated using the NEB method, combined with SCAN+\emph{U} calculations.
The computed migration barriers are  reported in Section 1 of the Supplementary Information (SI). We did not evaluate the migration energy of \ch{Na^+} at \ch{x}~=~2 in \NxTP{} and \NxCP{}, as well as composition \ch{x}~=~4 in {\NxCP{}}, since they have not been reported experimentally. 
 
The computed migration barrier (\Ebarrier) results are in good agreement with prior computational/experimental studies (see Section 1 of SI).  For example, for Na$_3$Cr$_2$(PO$_4$)$_3$ the \Ebarrier~$\sim$621~meV, agrees well with the $\sim$620~meV experimental value for $\alpha$-phase of the Na$_3$Cr$_2$(PO$_4$)$_3$ system.\cite{nogai_dipole_2018} Similarly, for Na$_3$V$_2$(PO$_4$)$_3$ we predict an \Ebarrier~$\sim$455~meV, which is comparable with the range of values 353--513~meV computed with the HSE06 functional.\cite{bui_hybrid_2015} In the case of \ch{Na_3Ti_2(PO_4)_3}, the predicted $E_{\rm barrier}$ $\sim$530~meV underestimates the experimental value of $\sim$750 meV from Ref.~\citenum{kabbour_-na_2011}. 

The directional dependence of \Ebarrier{} is removed by using the kinetically resolved activation barriers (\EKRA), as defined in Ref.~\citenum{van_der_ven_first-principles_2001}. Low values of \EKRA{}  correspond to low migration barriers and vice versa. The computed \EKRA{} values of \ch{Na^+} migration events of the type \ch{Na}(2)$\longleftrightarrow$\ch{Na}(1) of a MU that best represents the ground state configurations at different x are shown in Figure~\ref{fig:Structure_Ekra}(b). The lowest values of \EKRA{} are calculated at compositions \ch{Na_3M_2(PO_4)_3} and follow the order \ch{Ti} $\sim$282~meV $<$ \ch{V} $\sim$303~meV $<$ \ch{Cr} $\sim$348~meV. In contrast, the maximum \EKRA{} values are observed at \ch{x}$\sim$1, in agreement with existing reports.\cite{ishado_exploring_2020,nogai_dipole_2018,kabbour_-na_2011} The barriers at $x = 3$ across the different NaSICONs follow the ligand field stabilization energies (LFSE) for Ti$^{\rm III}(d^1)$~$<$~V$^{\rm III}(d^2)$~$<$~Cr$^{\rm III} (d^3)$, and the decreasing order of the transition metal sizes.{\cite{mckee_practical_1981,Cox1995}}  Note that LFSE also correlates with the NaSICON  pseudo-binary formation energies {\ch{Na_3Cr_2(PO_4)_3}}~$>>$~{\ch{Na_3V_2(PO_4)_3}} $>$ {\ch{Na_3Ti_2(PO_4)_3}}, thus increasing the Na migration barriers in the same order.\cite{singh_chemical_2021}  
  
Starting from the computed \EKRA{} encompassing several  Na-vacancy arrangements in the MU (see Section 2 of SI), a local cluster expansion (LCE) Hamiltonian\cite{van_der_ven_first-principles_2001} was trained for each  NaSICON system. We use the LCE together with our kinetic Monte Carlo simulation package\cite{deng_kmcpy_2022-1} to investigate the \ch{Na^+} transport within the NaSICON structures (see Sections 2 and 3 in the SI), by performing long-time (in the order of ms) and large-scale (8$\times$8$\times$8 formula units corresponding to 4096 Na sites) simulations. The LCE Hamiltonian could accurately predict the kinetic properties of {\ce{Na^+}} benchmarked on barriers obtained from first-principles calculations. The accuracy of predicted {\EKRA}s, obtained from the LCE formalism is bound within RMS errors of ${\pm}$31.75~meV for {\NxVP{}}, ${\pm}$25.29~meV for {\NxTP{}}, and ${\pm}$35.81~meV  for {\NxCP{}} (see Section 2 in the SI), which are within the perceived accuracy of migration barriers computed from first-principles ($\pm$50~meV ).{\cite{rong_materials_2015}}

By tracking all possible \ch{Na} migration events of each NaSICON, we simulated the \ch{Na^+} diffusion, quantified by: (\emph{i}) the jump diffusivity (D$_{\rm J}$), (\emph{ii}) the tracer diffusivity (D$^*$), and (\emph{iii}) the chemical diffusivity (D$_{\rm C}$).\cite{van_der_ven_rechargeable_2020} From the temperature vs.\ composition phase diagrams of these NaSICON systems,\cite{wang_phase_2022} using canonical Monte Carlo simulations, 1,850 initial configurations for each system
(50 different configurations at 37 unique Na compositions) with specific Na-vacancy arrangements were generated at $\sim$973~K, which is the typical synthesis temperature of these NaSICONs. Thus, these model structures mimic Na-vacancy configurations that are obtained post-synthesis,\cite{hong_crystal_1976,goodenough_fast_1976,park_crystal_2022,ishado_exploring_2020,kawai_high-voltage_2018,delmas_nasicon-type_1987} and are used in the kMC simulations as starting configurations. Subsequently, we performed 500 equilibration sweeps --one sweep is the total number of Na-vacancy sites in the simulation model, i.e., 4096-- followed by 3,000 kMC sampling sweeps of each configuration and statistically averaged the transport properties over a wide temperature range (300~K to 900~K, see Section 3 of the SI).
  
\begin{figure*}[!ht]
    \centering
    \includegraphics[width=\textwidth]{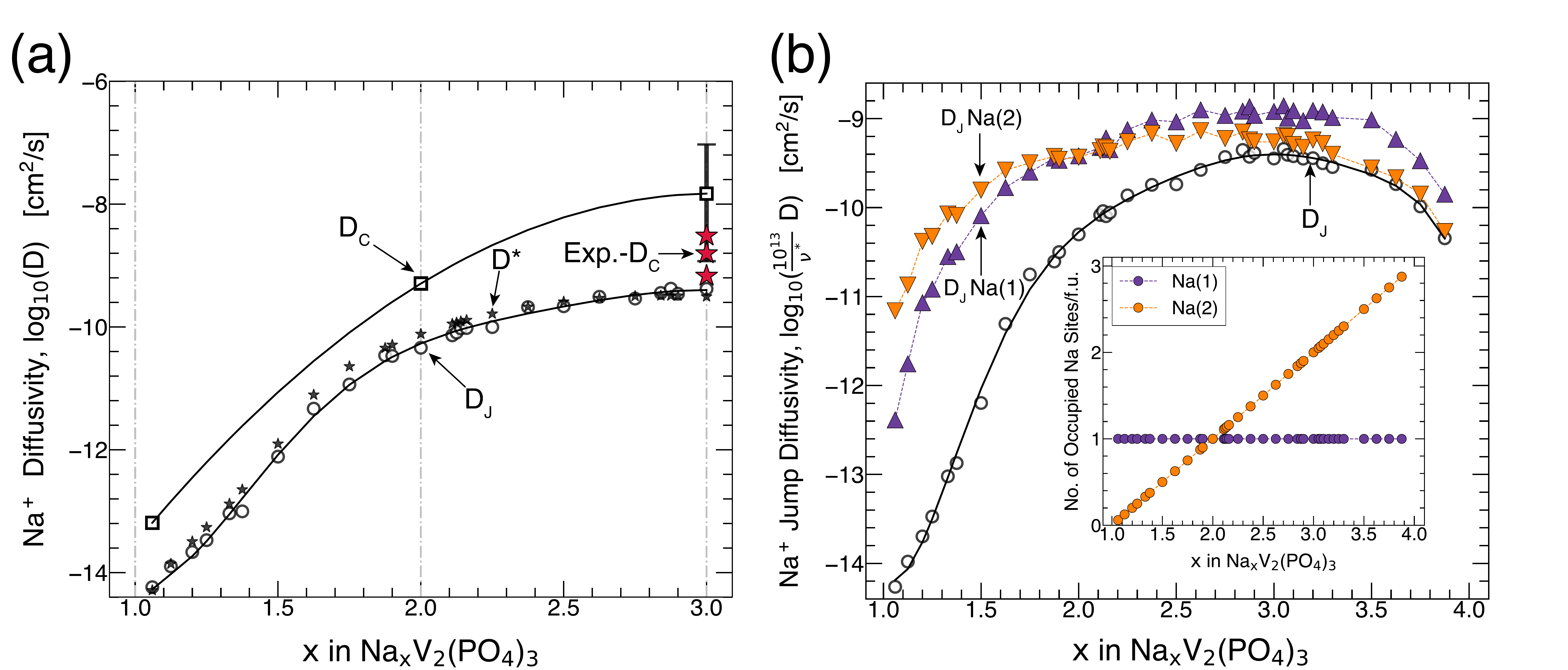}
    \caption{Predicted diffusivities of \NxVP{} at 300 K. Panel a plots D$_{\rm J}$ (circles), D$^*$ (stars), and D$_{\rm C}$ (squares). The pre-exponential factor is assumed as $1 \times 10^{13}$~Hz.\cite{van_der_ven_first-principles_2001} Vertical lines represent the phase boundaries derived from the phase diagram at $\sim$300~K.\cite{wang_phase_2022}  Values of D$_{\rm C}$ are only in the single-phase regions, i.e.\ \ch{x}~=~1, 2, and 3. At intermediate compositions  (two-phase regions) the thermodynamic factor $\Theta=0$ leads to zero D$_{\rm C}$. In the two-phase regions D$_{\rm C}$ is approximated using Vegard's law. The standard deviation of our predictions is shown at \ch{x}~=~3.0. Solid lines in black are the polynomial model fitted on the predictions.
    The experimental values of  chemical diffusivity (Exp.-D$_{\rm C}$, red stars) are  from Ref.~\citenum{fang_porous_2016} at x~=~3. Panel b plots the D$_{\rm J}$ (black circles), D$_{\rm J}$Na(1), which arises from the Na(1)-ion movement (purple triangles), and D$_{\rm J}$Na(2) from the Na(2)-ion movement (orange triangles). In panel b, all D$_{\rm J}$s are renormalized with a $\frac{10^{13}}{\nu^*}$ factor, due to the uncertainty in the prefactor $\nu^*$.\cite{kaxiras_adatom_1994,van_der_ven_first-principles_2001} 
    The inset shows the computed occupation number of Na(1) (in purple), and Na(2) (in yellow) sites per formula unit \textit{vs}.\ x extracted from the kMC simulations at 300 K. 
    }
    \label{fig:DAll_NVP}
\end{figure*}

We use \NxVP{} to discuss the behavior of the diffusion coefficients predicted by our kMC simulations. Figure~\ref{fig:DAll_NVP}(a) shows the computed D$_{\rm J}$, D$^*$, and D$_{\rm C}$ for \NxVP{} at 300~K as a function of \ch{x}. 

 We derived the D$_{\rm J}$ by tracking the center-of-mass of all the migrating \ch{Na^+} species, including the cross-correlations between different Na ions, which are excluded in the tracer diffusivity D$^*$. For this reason, in Figure~\ref{fig:DAll_NVP}(a), D$_{\rm J}$ and D$^*$ are different, but of similar magnitude, highlighting minimal contributions from cross-correlations, similar to observations in other electrode materials. \cite{van_der_ven_first-principles_2001} From the statistical analysis of the computed diffusivities we derived a standard deviation (Figure~\ref{fig:DAll_NVP}(a))  of approximately $\pm$2 orders of magnitude
in diffusivity  ($\pm120$~meV in terms of $E_{\rm barrier}$).\cite{rong_materials_2015} 

In \NxVP{} the \ch{Na^+} jump diffusivities of Figure~{\ref{fig:DAll_NVP}}(a) increase progressively from low \ch{Na} content ($\sim$5.77 $\times$ 10$^{-15}$ cm$^2$~s$^{-1}$ at \ch{x}~$\sim$~1) to high Na content ($\sim$4.16 $\times$ 10$^{-10}$ cm$^2$~s$^{-1}$ at \ch{x}~=~3.0).  Diffusivities for {\ch{Na_3V_2(PO_4)_3}} are in good agreement with existing measurements.\cite{ishado_exploring_2020,lan_ionic_2021,chen_challenges_2017} 

We derive a fourth order polynomial fit for the monotonically increasing D$_{\rm J}$ across 1~$\leq \rm x \leq~$3, namely $D_{\rm J} (x) = C + E_1 x + E_2 x^2 + E_3 x^3 + E_4 x^4$ (black line in Figure~\ref{fig:DAll_NVP}). The coefficients of the fit $E_1, \, E_2, \, E_3, \, {\rm and} \; E_4$ are reported in the SI. The fitted polynomial reflects the concentration dependence of  D$_{\rm J}$.

The chemical diffusivity D$_\mathrm{C}$ depends on  the thermodynamic factor $\Theta$, as per $\mathrm{D_{C}} = \mathrm{D_{J}} \Theta$; values of $\Theta$ are from Ref.~\citenum{wang_phase_2022}.  D$_{\rm J}$ and $\Theta$ contribute oppositely to D$_{\rm C}$. As $\Theta$ is related to the gradient of the Na chemical potential (Eq.\ 13 in SI), $\Theta$ usually takes large values for highly ordered (stable) configurations (Supporting Figure 16 in SI),
increasing D$_{\rm C}$ of the corresponding ordered phase. In contrast, low values of D$_{\rm J}$ are typically found in ordered phases, due to a lack of accessible vacant sites controlled by strong ion-vacancy ordering interactions. In \NxVP{}, D$_{\rm C}$ is  dominated by D$_{\rm J}$, as denoted by the relatively low  values of room temperature  chemical diffusivity at \ch{x}~=~1, and 2  (instead of the large values as controlled by $\Theta$).\cite{van_der_ven_first-principles_2001} 
Gray vertical lines in Figure~{\ref{fig:DAll_NVP}}(a) depict the phase boundaries  for \NxVP{} from Ref.~\citenum{wang_phase_2022} at 300~K. The predicted values of D$_{\rm C}$ at \ch{Na_3V_2(PO_4)_3} (1.18~$\times$~10$^{-9}$ -- 9.35~$\times$~10$^{-8}$ cm$^2$~s$^{-1}$) are higher than the experimental values (4.59~$\times$~10$^{-10}$ -- 2.0~$\times$~10$^{-9}$ cm$^2$~s$^{-1}$) measured by electrochemical impedance spectroscopy (EIS).\cite{fang_porous_2016,chen_challenges_2017}

While some experimental values of D$_{\rm C}$\cite{fang_porous_2016,chen_challenges_2017} fall within the standard deviations of our predictions, other experimental studies have reported significantly different D$_{\rm C}$ values, such as \num{3e-15} and \num{6e-13} cm$^2$~s$^{-1}$ measured using galvanostatic intermittent titration technique (GITT),\cite{bockenfeld_determination_2014} and \num{4e-14}--\num{2.48e-13} cm$^2$~s$^{-1}$ measured by EIS.\cite{shen_improved_2015} Such large differences in experimental D$_{\rm C}$ values can be attributed to different synthesis procedures of \ch{Na_3V_2(PO_4)_3}, which can affect the (im)purity of the particles, the particle sizes, and the defect concentrations, thus causing significant variations in transport properties. \cite{chen_challenges_2017} Nevertheless, we expect our kMC simulations to yield an accurate value of D$_{\rm C}$ within bulk \NxVP{} given that our confidence interval of calculated D$_{\rm C}$ is quite narrow.

The computed D$_{\rm C}$ for \NxVP{} at 300 K, 700 K, and 900 K (Supporting Figure 17 of the SI) is superimposed on the temperature-composition phase diagram, \cite{wang_phase_2022} where the single-phase regions are always connected by dashed lines representing the two-phase regions. For all temperatures explored, we observe an increase in D$_{\rm C}$ in the composition range  $1\le \rm x\le3$. Unsurprisingly, the D$_{\rm C}$ values increase for increasing temperatures, (i.e., at x~=~3 from \num{1.48e-8}~cm$^2$~s$^{-1}$ at 300~K to \num{2.5e-3}~cm$^2$~s$^{-1}$ at 900~K), signifying high thermally-activated motion of Na$^+$. 
 
Given the dominating contributions of D$_{\rm J}$ to the effective chemical diffusivity,\cite{van_der_ven_rechargeable_2020} we can gain valuable insights about the jump diffusivity (D$_{\rm J}$) of \NxVP{} at 300 K by separating D$_{\rm J}$ into the specific contributions from each sodium site, i.e., Na(1) and Na(2) in Figure~{\ref{fig:DAll_NVP}}(b). Here, we obtain D$_{\rm J}$Na(1) in the \NxVP{} structure by tracking all the possible \ch{Na^+} hopping events of the type Na(1)$\longrightarrow$Na(2), while we track Na(2)$\longrightarrow$Na(1) for D$_{\rm J}$Na(2). The differences between the overall D$_{\rm J}$ and individual D$_{\rm J}$\ch{Na}(1) and D$_{\rm J}$\ch{Na}(2) originate mainly from the number of the Na(1) and Na(2) sites available (see Eq.\ 8, 9, and 10 in SI). 

Values of D$_{\rm J}$Na(2) and D$_{\rm J}$Na(1) show similiar magnitudes,  and  appear higher than the overall D$_{\rm J}$, especially in the composition range  $1\le \rm x \le 3$. Both the D$_{\rm J}$Na(1) and D$_{\rm J}$Na(2) achieve their maxima at the intermediate Na compositions, specifically, \num{1.21e-9}~cm$^2$~s$^{-1}$ at x~$\approx$~2.94 for D$_{\rm J}$Na(1), and \num{1.15e-9} cm$^2$~s$^{-1}$ at x~$\approx$~2.71 for D$_{\rm J}$Na(2). We observe a sharp decrease of Na diffusivity near \ch{x} $\sim$ 1, where the values D$_{\rm J}$Na(2) and D$_{\rm J}$Na(1) are \num{6.91e-12}~cm$^2$~s$^{-1}$, and \num{4.11e-13} cm$^2$~s$^{-1}$, respectively.  

The number of occupied Na sites extracted from kMC simulations  (inset of Figure~\ref{fig:DAll_NVP}(b)) agrees well with existing site occupations at 300~K from experiment and theory.\cite{wang_phase_2022,park_crystal_2022} \ch{Na}(1) sites (purple line) are always fully occupied in the range 1 $\le \rm x \le$ 3, whereas the \ch{Na}(2) occupation (orange) varies from empty at \ch{x}~=~1 to fully occupied, with three \ch{Na}(2) sites per f.u., at \ch{x}~=~4. The number of occupied \ch{Na}(2) sites is much lower than that of \ch{Na}(1) at compositions around \ch{x}~=~1,\cite{wang_phase_2022} reflecting the difference in stability of the two sites at \ch{x}~=~1 which results in boosting the D$_{\rm J}$Na(2) over D$_{\rm J}$Na(1). Effects of cross-correlations among Na ions distributed between  \ch{Na}(1) and \ch{Na}(2) sites remain convoluted in the computed values of D$_{\rm J}$.

\begin{figure*}[!ht]
    \centering
    \includegraphics[width=\textwidth]{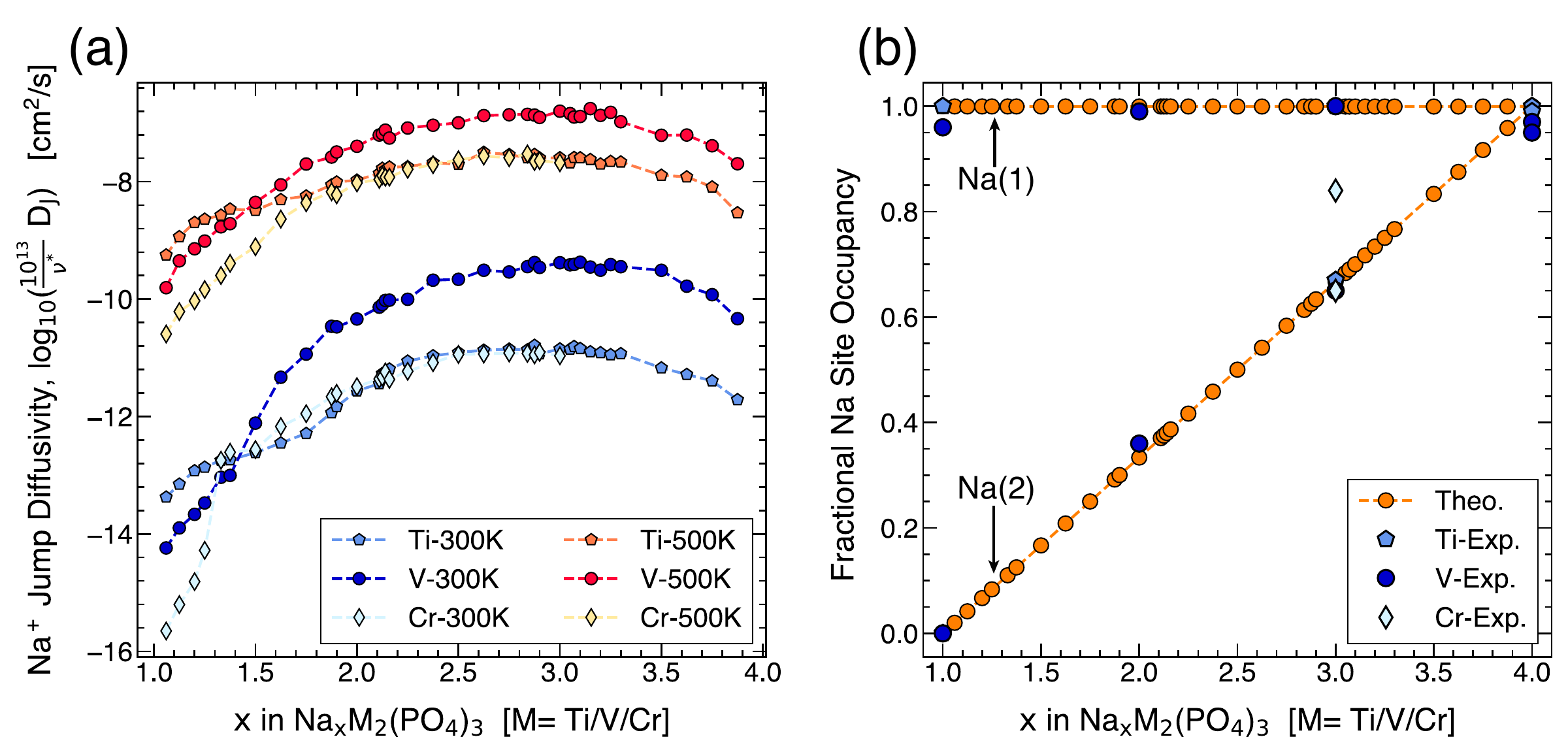}
    \caption{Computed Na$^+$ D$_{\rm J}$ at 300~K and 500~K (panel a) and fractional occupancy at 300~K (panel b) of Na(1) and Na(2) sites in Na$_{\rm x}$M$_2$(PO$_4$)$_3$ (M = Ti, V, or Cr). Due to the uncertainty in the prefactor $\nu^*$, all D$_{\rm J}$ values are normalized using a $\frac{10^{13}}{\nu^*}$.
    In panel a, data at 300~K, and 500~K are shown in blue and red/yellow symbols. For \NxCP{} Na extraction can only occur for $1 \leq \rm x  \leq 3$.\cite{kawai_high-voltage_2018} The computed results (Theo.) in panel b are denoted by orange circles, with experimental values in blue shapes.\cite{pang_synthesis_2014,senguttuvan_low-potential_2013,park_crystal_2022,chotard_discovery_2015,vijayan_physical_2011} The computed Na-site occupancy do not show significant differences among the three NaSICONs within our predictions, hence data for \NxVP{} is plotted in panel b. 
    }
    \label{fig:VTC_DJ_OCC}
\end{figure*}

We discuss the kMC predictions of two additional NaSICONs, \NxTP{}, and \NxCP{}, with similar electrochemical behavior to \NxVP{}.\cite{singh_chemical_2021,wang_phase_2022,chotard_discovery_2015,nogai_dipole_2018,kawai_high-voltage_2018,senguttuvan_low-potential_2013}  For  \NxTP{},  the sodium composition range $1 \leq \rm x \leq 3$ and $3 \leq \rm x \leq 4$ are accessible electrochemically,\cite{senguttuvan_low-potential_2013, singh_chemical_2021} whereas for \NxCP{}, \ch{Na} extraction only occurs in $1 \leq \rm x \leq 3$.\cite{kawai_high-voltage_2018,singh_chemical_2021} Similar to \NxVP{},  \NxTP{} and \NxCP{} show single phases at \ch{x}~=~1 and \ch{x}~=~3.\cite{pang_synthesis_2014,senguttuvan_low-potential_2013,vijayan_physical_2011,kawai_high-voltage_2018,singh_chemical_2021} Figure~\ref{fig:VTC_DJ_OCC}(a) shows the computed D$_{\rm J}$ of \NxVP{}, \NxTP{}, and \NxCP{} at 300 K and 500 K.

For all systems, we observe an increase of D$_{\rm J}$ in the composition range $1 \leq \rm x \leq 3$, followed by a gradual decrease of  D$_{\rm J}$ for $3 < \rm x \leq 4$ in \NxTP{} and \NxVP{}. At 300 K, D$_{\rm J}$ reaches a maximum for all systems at \ch{x}$\sim$3, with \NxVP{} displaying a higher magnitude of D$_{\rm J}$ in the range 1.5 $\leq \rm x \leq$ 4 than \NxTP{} and \NxCP{}. At 500 K, in comparison, the D$_{\rm J}$ of all the three NaSICONs were of a similar order of magnitude, corresponding intuitively to higher diffusion rates with increasing temperature. 

The maximum values of D$_{\rm J}$ occur at similar compositions for all NaSICONs, specifically \num{4.09e-10}~cm$^2$~s$^{-1}$ for \NxVP{} at \ch{x}~$\sim$~2.9, \num{1.71e-11}~cm$^2$~s$^{-1}$ for \NxTP{} at \ch{x}~$\sim$~2.7, and \num{1.39e-11}~cm$^2$~s$^{-1}$ for \NxCP{} at x~$\sim$~2.6, respectively. Similarly, the lowest values of D$_{\rm J}$ occur at low Na content (i.e., \ch{x} $\sim$ 1) for all NaSICONs. 
For example, the lowest D$_{\rm J}$ among the NaSICONs at x~$\sim$~1 is \num{2.2e-16}~cm$^2$~s$^{-1}$ for \ch{Na_1Cr_2(PO_4)_3}. The low values of D$_{\rm J}$ at x~$\sim$~1 suggest that the reversible extraction of the ``last'' Na-ion from NaSICON electrodes may be limited also by the poor kinetics at low Na content. 

To understand the Na$^+$ distributions at 300 K, we extract the fractional occupancies of Na(1) and Na(2) sites (Figure~\ref{fig:VTC_DJ_OCC}(b)), and compare them with the experimental data of these NaSICON systems. At room temperature, we observe high stability of the Na(1) site across the entire Na concentration range ($1\leq \rm x \leq 4$), where it remains fully occupied, while Na(2) occupancy monotonically increases with increasing \ch{x}. Our kMC results agree quantitatively with the experimentally refined Na(1)/Na(2) occupations.\cite{pang_synthesis_2014,senguttuvan_low-potential_2013,park_crystal_2022,chotard_discovery_2015,vijayan_physical_2011} 
 In the case of Na$_3$Cr$_2$(PO$_4$)$_3$ the predicted occupation of Na(1) ($\sim$1) overestimates the experimental data ($\sim$0.84), which suggests that the nominal composition may deviate from the real composition of the synthesized NaSICON.\cite{vijayan_physical_2011}

\section{Discussion} 

In this letter, using kMC simulations that bear the accuracy of DFT calculations, we investigated the variability of Na-ion transport vs.\ Na content in three important NaSICON electrodes for Na-ion batteries: \NxTP{}, \NxVP{}, and \NxCP{}. We demonstrated that the ion transport properties (D$_{\rm J}$, D$^*$, and D$_{\rm C}$) of NaSICONs have a configurational dependence on the local sodium vacancy arrangements near the migration events. 

In the three NaSICONs, we observed an increase of D$_{\rm J}$ from low Na concentrations of x~$\sim$~1 to x~$\sim$~3. For \NxVP{} and \NxTP{}, we observed a decrease in D$_{\rm J}$ in the composition range $3 < \rm x \leq 4$ (Figure~\ref{fig:VTC_DJ_OCC}(a)). \break

\noindent {\bf Interplay of Ligand Field Stabilization and Ionic Radii of Transition Metals:} Two main factors: (\emph{i}) the ligand field stabilization of transition metals,  and their (\emph{ii}) ionic radii can explain variations of migration barriers and diffusivities in these NaSICONs as a function of Na composition.

Different NaSICONs exhibit different migration barriers at the same \ch{Na} content (see Figure~\ref{fig:Structure_Ekra}(b)), with implications on the observed D$_{\rm J}$. For example, at x~$\sim$~1, the highest \EKRA{} $\sim$591~meV is exhibited by \ch{Na_1Cr_2(PO_4)_3}, well exceeding  447~meV in \ch{Na_1V_2(PO_4)_3}, and 371~meV in \ch{Na_1Ti_2(PO_4)_3}. Since at x~$\sim$~1 all transition metals, Cr, Ti, and V are tetra-valent (verified by the magnetic moments of Supporting Figure 7 in SI), differences in migration energies in NaSICONs at this composition are controlled by the transition metal ionic radii, which will lead to different sizes of ``bottlenecks'' for the migrating Na-ion. \cite{park_crystal_2022}

At x = 1, the transition metal ionic radii follow the trend of Ti$^{\rm IV}$ ($\sim$0.61~\AA)~$>$  V$^{\rm IV}$ ($\sim$0.58~\AA) $>$ Cr$^{\rm IV}$ ($\sim$0.55~\AA), which causes a similar variation in the lattice parameters. \cite{singh_chemical_2021, park_crystal_2022, pang_synthesis_2014} Thus, the sizes of bottleneck (i.e., the transition state) for the migrating Na$^+$ reduce in the order of \ch{Na_1Ti_2(PO_4)_3} $>$ \ch{Na_1V_2(PO_4)_3} $>$ \ch{Na_1Cr_2(PO_4)_3},\cite{park_crystal_2022} reflecting an identical trend in D$_{\rm J}$ values observed (see Figure~{\ref{fig:VTC_DJ_OCC}}(a)). Furthermore, the polyhedral volumes of  transition-metal octahedra are $\sim$9.80~\AA$^{3}$ for {\ch{TiO_6}}, $\sim$9.33~\AA$^{3}$ for {\ch{VO_6}}, and $\sim$8.77~\AA$^{3}$ for {\ch{CrO_6}}. The bond distance between the migrating Na$^{+}$ at the transition state and nearby O$^{2-}$ (which appears linearly correlated with the migration barriers{\cite{song_first_2014, guin_new_2016}}) follows a sequence of 
{\ch{Na_1Ti_2(PO_4)_3}} ($\sim$2.36~\AA) $>$ {\ch{Na_1V_2(PO_4)_3}} ($\sim$2.30~\AA) $>$ {\ch{Na_1Cr_2(PO_4)_3}} ($\sim$2.27~\AA). This sequence  indicates the narrower sizes of the migration ``bottleneck'' from {\ch{Na_1Ti_2(PO_4)_3}}, via {\ch{Na_1V_2(PO_4)_3}}, to {\ch{Na_1Cr_2(PO_4)_3}}, hence explaining the reduced values of D$_{\rm J}$ in Figure~{\ref{fig:VTC_DJ_OCC}}(a).

The impact of transition metal ionic radii is lower at higher Na content (i.e., at x~$\sim$~3), since Na$^+$ dictates lattice parameters more significantly, given its large ionic size. Thus, variations observed in \Ebarrier{} and $D_{\rm J}$ can be attributed more to the electronic structure (and LFSE) of the transition metals involved rather than changes in lattice parameters or bottleneck sizes. Specifically, LFSE stabilizes Cr$^{\rm III}$ ($3d^3$) more significantly than V$^{\rm III}$ ($3d^2$) (see Supporting Figure 8 in SI), due to the high stability of the half-filled high-spin $t_{2g}$ orbitals of Cr$^{\rm III}$.\cite{singh_chemical_2021} A higher Na$^+$ \Ebarrier{} in \ch{Na_3Cr_2(PO_4)_3} compared to \ch{Na_3V_2(PO_4)_3} is expected as more energy is required to oxidize Cr$^{\rm III}$ that is near a migrating Na$^+$ to Cr$^{\rm IV}$, which is consistent with a lower value of D$_{\rm J}$ (Figure~{\ref{fig:VTC_DJ_OCC}}(a)) for \ch{Na_3Cr_2(PO_4)_3} than \ch{Na_3V_2(PO_4)_3}. 

In the case of \ch{Na_3Ti_2(PO_4)_3}, Na$^+$ migration is penalised compared to \ch{Na_3V_2(PO_4)_3} since the highly stable configuration of Ti$^{\rm IV}$($3d^0$) needs to be reduced by the migrating Na$^+$ to the un-preferred Ti$^{\rm III}$($3d^1$) configuration. Thus, the energy cost associated with a local Ti reduction near a migrating Na causes the D$_{\rm J}$ for \ch{Na_3Ti_2(PO_4)_3} to be lower than \ch{Na_3V_2(PO_4)_3}. In general, at composition x~$>$~2 {\NxVP{}} shows consistently higher D$_{\rm J}$ compared to {\NxTP{}} and {\NxCP{}} (Figure~{\ref{fig:VTC_DJ_OCC}}(a)). \break

\noindent {\bf Extracting the last Na from \ch{\mathbf{Na_1M_2(PO_4)_3}}:} From the occupation of Na sites (Figure~\ref{fig:VTC_DJ_OCC}(b)) at x~=~1, we identified only Na(1) sites were fully occupied, whereas Na(2) sites were empty. Such an arrangement of the Na-ions relates to the ``structural integrity'' of \ch{Na_1M_2(PO_4)_3}, where only Na(1) is occupied and screens the electrostatic repulsions of nearby MO$_6$ octahedra stacked along the $c$-axis imparting stability to the Na$_1$M$_2$(PO$_4$)$_3$.\cite{park_crystal_2022,chen_nasicontype_2019,liu_exploring_2017,wang_highenergy_2020} 
Furthermore, we observed a large site energy difference ($\sim$880~meV) between the lower energy Na(1)  site and Na(2)  at x~$\sim$~1 in {\ch{Na_1V_2(PO_4)_3}} (see Supporting Figure 2 of SI). These results indicate that Na(2) sites are thermodynamically unstable at x~$\sim$~1, which hinders the ion transport in {\ch{Na_1M_2(PO_4)_3}}.\cite{deng_phase_2020} We also observed an abrupt decline of D$_{\rm J}$ (i.e., $\sim$\num{5.8e-15}~cm$^2$~s$^{-1}$) in \ch{Na_1V_2(PO_4)_3}, suggesting that further \ch{Na} extraction $x<1$ is impractical. Such a drop of  diffusivity at \ch{x}~=~1 has also been commented on in prior reports.\cite{chen_challenges_2017,lan_ionic_2021,bockenfeld_determination_2014,niu_detailed_2016}

Low values of D$_{\rm J}$ at low Na concentrations limit the full utilization of the NaSICON capacities. The extraction of the last Na-ion should happen via {\ch{Na_1V^{IV}V^{IV}(PO_4)_3}}~$\leftrightarrow$~{\ch{V^{V}V^{IV}(PO_4)_3}}~+~{\ch{e^-}}~+~{\ch{Na^+}}, in the case of \NxVP{}. This reaction is not redox-limited, as the high-potential $\rm V^{V}/V^{IV}$ redox couple is reversibly accessible.{\cite{lalere_improving_2015,wang_triggering_2020,wang_importance_2021}} To date, the chemical Na extraction from {\ch{Na_1V_2(PO_4)_3}} has been elusive.{\cite{gopalakrishnan_vanadium_1992}} The impractical extraction from {\ch{Na_1V_2(PO_4)_3}} can be partially attributed to the low Na-diffusivities (Figures~{\ref{fig:DAll_NVP}} and {\ref{fig:VTC_DJ_OCC}}).{\cite{chen_challenges_2017,bockenfeld_determination_2014,lan_ionic_2021}}  

 We propose that atomic substitutions on either Na sites, transition metal sites, or the inclusion of alternative polyanion groups may represent the practical approaches to facilitate the extraction of the last Na-ion in these NaSICONs. The feasibility of such a method has been confirmed in several studies.\cite{lalere_improving_2015,liu_exploring_2017,patoux_structural_2003,park_asymmetric_2022} The last Na-ion can be extracted from {\ch{Na_1Nb_2(PO_4)_3}} forming the mix-valence {\ch{Nb_2(PO_4)_3}},{\cite{patra_unveiling_2023}} or from  the mixed NaSICON {\ch{Na_xTiNb(PO_4)_3}}.{\cite{bennouna_specificites_1995,tillement_crystal_1991}} Thus, mixing {\ch{V}} with ``softer'' transition metals from the 2$^\mathrm{nd}$ and 3$^\mathrm{rd}$ rows, bearing similar redox characteristics to vanadium, may unlock additional capacity in NaSICONs. \break  

\begin{figure}[!ht]
    \centering
    \includegraphics[width=1.\columnwidth]{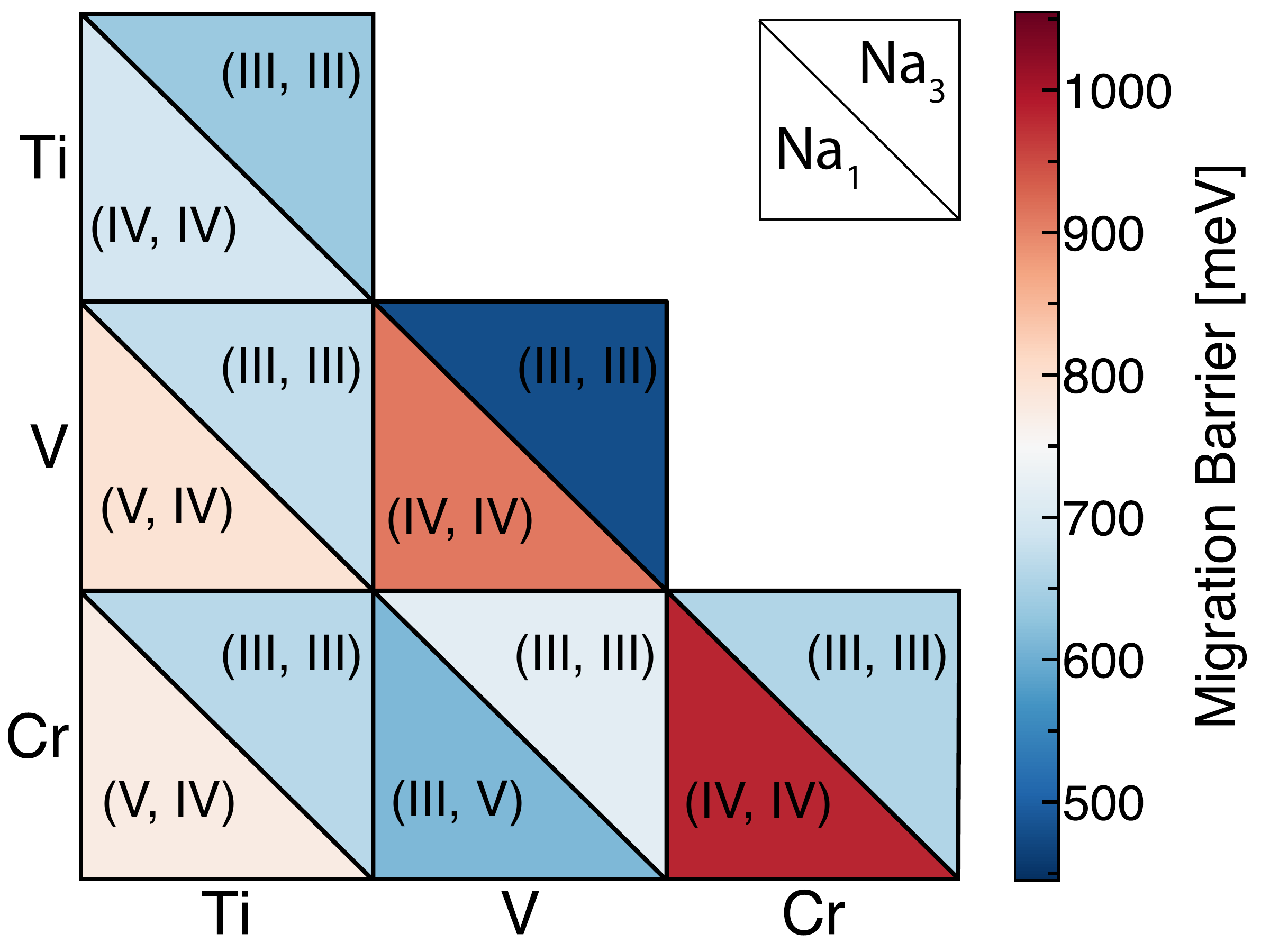}
    \caption{Migration barriers for \ch{Na_1MM'(PO_4)_3} and \ch{Na_3MM'(PO_4)_3} with mixed transition metal combinations, where M, M'~=~Ti, V, or Cr. The ratio of transition metals is kept at 1:1. The three diagonal squares are 1-transition-metal NaSICONs, where only consistent tetra-valent (IV), and tri-valent (III) states were observed at Na$_1$ and Na$_3$ compositions. Each square is divided into a lower triangle, and a higher triangle, corresponding to the migration barriers for Na$_1$, and Na$_3$ compositions, respectively. Within each triangle, we identify the oxidation states of the two transition metals, which provide the local charge ordering environment for migrating Na$^+$, as (m, n), where m and n correspond to the oxidation states of transition metal labeled on the y-axis, and x-axis, respectively. For example, (III, V) in the lower-triangle of \ch{Na_1CrV(PO_4)_3} square identifies the local charge arrangement of Cr$^{\rm III}$, and V$^{\rm V}$ oxidation states.
    }
    \label{fig:Eb_mixed}
\end{figure}

\noindent {\bf Migration barriers in mixed-transition-metal NaSICONs:} On the one hand, at low Na content (i.e., x~=~1) transition metals with higher oxidation states will tend to repel the Na-ions at Na(1) electrostatically, and hence  destabilize the Na(1) site with an increase of the site energy at Na(1). In addition, the occurrence of higher oxidation state transition metals will decrease the electron density on the surrounding \ch{O^{2-}}, which will reduce their electrostatic attraction to Na--ions and increase the repulsion between nearby ``O$_3$'' faces of the MO$_6$ octahedra, consequently enlarging the Na-migration bottlenecks.\cite{park_crystal_2022}  On the other hand, transition metals with a lower oxidation state will attract more Na-ions around, resulting in the population of nearby Na(2) sites, which may lower the site energy difference between Na(1) and Na(2) sites at x~=~1. 
These considerations suggest that local charge arrangement with higher/lower oxidation states of transition metals may introduce disorder on the Na-vacancy lattice, which may decrease the \Ebarrier{} at x~=~1. To quantify this aspect, we evaluated additional Na-migration barriers for  NaSICONs with mixed transition metals at a 1:1 ratio (see Figure~\ref{fig:Eb_mixed}). 

In Na$_1$VTi(PO$_4$)$_3$ and Na$_1$VCr(PO$_4$)$_3$, we observed \Ebarrier{} of $\sim$794~meV, and $\sim$613~meV, respectively (see Supporting Figure 10 in SI), lower than the $\sim$910~meV in Na$_1$V$_2$(PO$_4$)$_3$. Instead of the single oxidation state of V$^{\rm IV}$ observed at Na$_1$V$_2$(PO$_4$)$_3$, we found local charge arrangement of V$^{\rm V}$ and Ti$^{\rm IV}$ redox states near the migrating Na-ion in Na$_1$VTi(PO$_4$)$_3$. Similarly, in Na$_1$VCr(PO$_4$)$_3$, suggest an ordering of the  V$^{\rm V}$ and Cr$^{\rm III}$ states (see Supporting Figure 9 in SI) in agreement with existing experiments.\cite{wang_sodium_2017,liu_exploring_2017} Furthermore, the site energy difference between Na(2) and Na(1) also decreased from $\sim$880~meV for Na$_1$V$_2$(PO$_4$)$_3$ to 744~meV for Na$_1$VTi(PO$_4$)$_3$, and 576~meV for Na$_1$VCr(PO$_4$)$_3$.

The lower values of \Ebarrier{} for mixed transition-metal-NaSICONs at x = 1 will increase D$_{\rm J}$ for Na$_1$VTi(PO$_4$)$_3$, and Na$_1$VCr(PO$_4$)$_3$ by $\sim$2, and $\sim$4 orders of magnitude, respectively,\cite{rong_materials_2015} in comparison to D$_{\rm J}$ of Na$_1$V$_2$(PO$_4$)$_3$.  A lower \Ebarrier{} of 774~meV for Na$_1$TiCr(PO$_4$)$_3$ will also increase its D$_{\rm J}$ by approximately 3 orders of magnitude compared to Na$_1$Cr$_2$(PO$_4$)$_3$. 

In \ch{Na_3VTi(PO_4)_3}, \ch{Na_3VCr(PO_4)_3}, and \ch{Na_3TiCr(PO_4)_3}, the \Ebarrier{} are  $\sim$676~meV, $\sim$719~meV, and $\sim$667~meV, respectively (see Supporting Figure 12 in SI). Only trivalent transition metals, such as V$^{\rm III}$, Ti$^{\rm III}$, and Cr$^{\rm III}$ were found in these systems (see Supporting Figure 11 in SI), which set specific local charge arrangements for the migrating Na$^+$. Compared with the migration barriers for the Na$_3$V$_2$(PO$_4$)$_3$, Na$_3$Ti$_2$(PO$_4$)$_3$, and Na$_3$Cr$_2$(PO$_4$)$_3$ NaSICON analogs at x~=~3, which were $\sim$485~meV, $\sim$638~meV, and $\sim$659~meV, respectively, the deviations between \Ebarrier{} of single-transition-metal NaSICON and mixed-transition-metal NaSICON are as small as $\sim$50~meV, except Na$_3$V$_2$(PO$_4$)$_3$, which exhibits a significantly lower \Ebarrier{}.

Our results in Figure~{\ref{fig:Eb_mixed}} indicate that the local charge arrangements on the transition metal sites with higher/lower oxidation states may disrupt potentially stable Na--Va arrangements at Na$_1$M$_2$(PO$_4$)$_3$, thus lowering the migration barriers and enhancing the Na$^+$ D$_{\rm J}$, which, in turn, may enable the extraction of the last Na. For example, in the case of the  \NxVP{} system, if compositions with low Na content  (i.e., {\ch{Na_{0+x}V_2(PO_4)_3}}) were thermodynamically stable, the Na extraction from  {\ch{Na_1V_2(PO_4)_3}}  would be highly facilitated because of the favorable local charge arrangement of the mixed-valence vanadium sites V$^{\rm IV/V}$.

\section{Conclusion}
In conclusion, our \emph{ab initio}-based kMC approach revealed the complex relationships between Na-ion transport in \NxTP{}, \NxVP{}, and \NxCP{} NaSICON electrode materials as a function of Na content and temperatures. We identified optimal compositions providing maximum intrinsic Na$^+$ diffusivity for \NxTP{}, \NxVP, and \NxCP{}. Our analysis demonstrated that the Na transport properties of NaSICON materials are highly dependent on the local chemical environments determined by the local arrangements of sodium ions and their vacancies, as well as the oxidation states of transition metals. In particular, we elucidated that the environments favoring stable Na-vacancy orderings, typically in the fully charged region, should be disrupted to increase the energy density of NaSICON electrodes, such as \NxVP{}. The insights gained from this study into the Na$^+$ diffusion properties in \NxTP{}, \NxVP{}, and \NxCP{} shed light on appropriately tailored combinations of transition metals that can be used to access swift Na transport in polyanionic electrodes for inexpensive Na-ion batteries.

\begin{acknowledgement}
We acknowledge funding from the National Research Foundation, Singapore under his NRF Fellowship NRFF12-2020-0012. Z.\ D.\ acknowledges the support from his Lee Kuan Yew Postdoctoral Fellowship 22-5930-A0001. The computational work was performed on resources of the National Supercomputing Centre, Singapore (https://www.nscc. sg).
\end{acknowledgement}

\begin{suppinfo}
The Supporting Information is available free of charge and includes (\emph{i}) Details of simulations of migration barriers with first-principles methods, (\emph{ii}) The formalism of the local cluster expansion, and (\emph{iii}) Details of kinetic Monte Carlo simulations.
\end{suppinfo}

\bibliography{biblio}

\end{document}